\documentstyle[aps,twocolumn,epsf]{revtex}
\def\ave#1{\langle #1\rangle}

\newcommand{\op}[1]{\hat{#1}}

\newcommand{\bra}[1]{\langle #1|}
\newcommand{\ket}[1]{|#1\rangle}
\newcommand{\braket}[2]{\langle #1|#2\rangle}

\begin{document}
\title{Non ergodic quantum behaviour in classically chaotic 3D billiards
}
\author{Giulio Casati}
\address{International Center for the study of dynamical systems,
University of Milano at Como, Via Lucini, 3, 22100 Como, Italy
 Istituto Nazionale di Fisica della Materia and INFN sezione di Milano,
Via Celoria 16, 20133 Milano, Italy}
\author{Toma\v z Prosen}
\address{Physics Department, Faculty of Mathematics and Physics,
University of Ljubljana, Jadranska 19, 1000 Ljubljana, Slovenia
}
\date{\today}
\draft
\maketitle
\begin{abstract}
We study, analytically and numerically, the classical and quantum
properties of a nearly spherical 3D billiard. In particular we show the
appearence of quantum non ergodic behaviour and of the deviations from
Random Matrix Theory
predictions which are due to the quantum suppression of classically
chaotic diffusion.
\end{abstract}
\vfill

In the ergodic theory of classical dynamical systems, billiards have 
played a fundamental role since they exhibit a
clean and rich variety of dynamical properties, from completely integrable
to true random motion. For this reason, since the early papers
\cite{CGV80}
in which level spacing distribution was numerically studied, they
became of primary importance for the theoretical and experimental analysis
of the various quantum properties which are connected to different
features of classically chaotic motion.
An important question is to understand under what conditions the 
predictions of Random Matrix Theory (RMT) can be applied to a generic 
conservative quantum system\cite{BGS84}. In this context the quite
surprising
phenomenon of localization inside the energy shell has been shown to play
a fundamental role\cite{CCGI}. This phenomenon implies quantum
non-ergodic behaviour and deviations from the predictions of Random
Matrix theory. Moreover a very interesting asymmetry appears between
the structure of exact and unperturbed eigenfunctions.

In a recent paper \cite{BCL96} quantum localization was shown to take place
in a stadium billiard and this result was confirmed also in a similar
model \cite{FS97}. The important question is now what happens in more
dimensions. In particular we would like to understand  whether or not this
interesting non
ergodic quantum behaviour can take place in a  3D billiard. 3D billiards
 are certainly more complicated than plane
billiards. However, in some cases, they can be more close to experimental
investigations \cite{30}-\cite{33} and to 
physical applications; we refer for example to dielectric resonators
which may find applications in microlasers and fibre-optic communications
\cite{STO}
\\\\
In this paper we consider  a wiggly 3D spherical billiard, namely a
particle of unit mass
and velocity $\vec{v}$ moving freely inside a closed 3D domain and 
bouncing elastically off the boundary whose 
shape is given by a small, smooth but wiggly deformation of the unit
sphere.
The boundary of the billiard is described by the distance of the boundary
point from the origin as a function of the unit direction vector $\vec{n},\;
r_B(\vec{n}) = 1 + \epsilon f(\vec{n})$ ($\vec{n}^2 = 1$).
The small parameter $\epsilon$ controls the size of the wiggles of the boundary
while $f(\vec{n})$ is an oscillatory function
which is assumed to have unit average square modulus
$\ave{[f(\vec{n})]^2} = (1/4\pi)\int d^2\vec{n} [f(\vec{n})]^2 = 1$. 
The shape function $f(\vec{n})$ may be expanded in terms of spherical 
harmonics $Y_{\ell m}(\vec{n})$ with a given maximal order $L$,
$f(\vec{n}) = 
\sum_{\ell m}^{\ell \le L} c_{\ell m} Y_{\ell m}(\vec{n}).$
The coefficients $c_{\ell m}$ have been specified by imposing three
conditions:
(i) $c_{\ell m}$ decrease with increasing order $\ell $ as
$\sum_m c_{\ell m}^2 \propto 1/\ell^2$,
(ii) the shape of the billiard is invariant under the cubic symmetry
group $O_h$, $f(\vec{n}) = f(G\vec{n}),\; G\in O_h$,
(iii) the shape function $f(\vec{n})$ is separable in cartesian
coordinates,
i.e. it
can be written in a form 
$ f(\vec{n}) = \sum_{p=2}^{L/2} a_p (n_x^{2p} + n_y^{2p} + n_z^{2p})$ 
with coefficients $a_p$.
\\\\
By calculating the distribution of maximal Liapunov exponent,
the classical dynamics of the billiard has been found
to be
almost completely chaotic for $\epsilon > \epsilon_c(L)$, where the
critical chaos border $\epsilon_c(L)$ is a rapidly decreasing function of
$L$.
For the case  $L=14$ which will be taken in this paper,
we found $\epsilon_c=0.001$. 
For this value of the perturbation, most of phase space is 
covered by chaotic orbits. The billiard's dynamics can be described 
in terms of the angular momentum 
$\vec{l} = \vec{r}\times\vec{v}$ 
and the unit direction vector $\vec{n} = \vec{r}/r$ of the forthcoming point 
of collision with the boundary.
Labelling points of successive collisions by the discrete time variable
$j$,
one can write an approximate Poincare map for the dynamics in 
$(\vec{l},\vec{n})$ variables:
\begin{eqnarray}
\vec{l}_{j+1} &=& \vec{l}_j - 2\epsilon\left[(v^2-l_j^2)^{1/2}
\vec{n}_j\times\nabla f(\vec{n}_j)\right]
+ {\cal O}(\epsilon^2),\label{eq:map}\\
\vec{n}_{j+1} &=& (2l^2_{j+1}/v^2 - 1)\vec{n}_j -
2(v^2-l^2_{j+1})^{1/2}\vec{n}_j\times\vec{l}_{j+1}/v^2, \nonumber
\end{eqnarray}
which is essentially four dimensional since we have two constraints, namely
$\vec{l}_j\cdot\vec{n}_j = 0$ and $\vec{n}_j^2 = 1$. The components of $\vec{l}_j$ are 
the {\em slow} or adiabatic variables while $\vec{n}_j$ is the {\em fast} variable (only 
one of its components is independent due to constraints). 
By averaging over the fast variable one can calculate 
the average drift of  the angular momentum vector
$\Delta\vec{l}_j=\vec{l}_{j+1}-\vec{l}_j$,
\begin{eqnarray}
\ave{\Delta\vec{l}_j} &=& \vec{\Omega}(\vec{l}_j)\times\vec{l}_j,
\label{eq:drift} \\
\vec{\Omega}(\vec{l}) &=& 2\epsilon\frac{\sqrt{v^2-l^2}}{l^2}
\sum_{p=1}^{L/2-1}
\frac{(2p+1)!!}{(2p)!!} a_{p+1} \nonumber\\
&\times&\left(
\frac{l_x l^{2p}}{(l^2-l_x^2)^p},
\frac{l_y l^{2p}}{(l^2-l_y^2)^p},
\frac{l_z l^{2p}}{(l^2-l_z^2)^p}\right)\nonumber
\end{eqnarray}
From eq. (\ref{eq:drift}) it is seen
that the direction of angular momentum vector undergoes a 
precession with a frequency vector field 
$\vec{\Omega}(\vec{l})={\cal O}(\epsilon L^{1/2})$
which is an oscillatory function of angular momentum on a scale
$v/L$.

Let us first analyze the behaviour of the modulus $l$ of angular momentum.
From (\ref{eq:drift}) we see that the average drift of $l$ is zero 
$\ave{\Delta l_j} = \ave{l_{j+1}-l_j} \approx 
(\vec{l}_j/l_j)\cdot\ave{\Delta\vec{l}_j} = 0.$
Then, above the chaos border $\epsilon > \epsilon_c$,
one typically expects the diffusive behaviour  $\ave{(l_J-l_0)^2} \approx
D(l_0) J $
where the diffusion coefficient can be estimated from (\ref{eq:drift})
as $D = {\cal O}(\epsilon^2 v^2 L^2)$.
However, our numerical analysis reveals that the above diffusive behaviour
takes place only up to time $J = J^*$ where $J^* \approx L$. 
Instead, for $J > J^*$, we find a slower, sublinear diffusion, 
which obeys the empirical law $\ave{(l_J-l_0)^2} \propto J^\alpha$ with 
$0 < \alpha < 1$ \cite{expl}.

For the variance of the component $l_z$ of angular momentum, both,
the drift and diffusive terms are important
$\ave{(l_{zJ}-l_{z0})^2} \approx 
\ave{l_{zJ}-l_{z0}}^2 +  J (\ave{(\Delta l_{z0})^2} - \ave{\Delta l_{z0}}^2)\\ 
\approx J^2 (\vec{\Omega}\times\vec{l}_0)_z^2 + J D(l_0).$
The first (drift) term is ${\cal O}(J^2\epsilon^2 v^2 L)$ 
and the second (diffusive) term is
${\cal O}(J\epsilon^2 v^2 L^2)$, so the drift may be neglected w.r.t. 
diffusion for $J < J^*\approx L$, while for $J > J^*$ the quadratical 
drift term prevails \cite{100}.
 It is not the purpose of the present paper to discuss the details of
the classical diffusive mechanism. We would like only to stress that, as
confirmed by our numerical analysis, the
`diffusion' of the direction of angular momentum is much faster than the 
diffusion of its modulus which is therefore the slowest variable in the
system.\\\\
The classical steady state equilibrium distribution $\rho_{\rm eq}(l)$ of
the
angular momentum can be obtained analytically by substituting
(provided $\epsilon > \epsilon_c$)
time averages with phase averages:
\begin{equation}
\rho_{\rm eq}(l) = \lim_{J\rightarrow\infty}\frac{1}{T}
\sum\limits_{j=1}^J\Delta t_j\delta(l-l_j) = \frac{3l}{v^3}\sqrt{v^2-l^2}
\label{eq:ssd}
\end{equation}
Here $\Delta t_j = 2(v^2-l_j^2)^{1/2}/v^2$ are the time 
intervals between collisions.
As a measure of the width of the classical distribution we introduce the
quantity
$\sigma(l_0,t) = (1/v)\sqrt{\int_0^v dl\rho(l,t)l^2 - 
(\int_0^v dl\rho(l,t)l)^2}$
with  $\rho(l,t=0) = \delta(l-l_0)$.
For the equilibrium steady state we find $\sigma_{\rm eq} = \sigma(l_0,
\infty) = 0.2303$. We have used espression (\ref{eq:ssd}) as an additional
numerical test of ergodicity
of our billiard for $\epsilon > \epsilon_c$.

Let us now turn to the quantum dynamics.
We have solved the Schr\" odinger equation 
$(\nabla^2+k^2)\Psi_k(\vec{r})=0$ with Dirichlet boundary
conditions ($\hbar=1, E=k^2/2$).
Eigenvalues $k^{(n)}$ and eigenfunctions $\Psi_{k^{(n)}}$ have been 
computed very efficiently using the scaling method invented by 
Vergini and Saraceno \cite{VS95}
and already implemented for smooth 3D billiards in \cite{P96}.
Note that we have studied only the states belonging to 1-dim
fully symmetric irrep of the 48-fold cubic group $O_h$, so the 
Weyl formula which relates the wavenumber $k$ with the sequential quantum
number ${\cal N}$ in this symmetry class reads  (for small $\epsilon$) ${\cal N}(k) \approx k^3/(216\pi)$.
For sufficiently large  $k$ we expect that the statistical
properties of eigenvalues
and eigenfunctions of our billiard are well described by RMT.
On the other hand, for sufficiently small $k$ (or small $\epsilon$ and
fixed 
$k$) the perturbation 
theory is adequate and one can treat the problem as a small perturbation of a 
spherical billiard.
We estimate the perturbative border $\epsilon_p(k)$ as the critical
value at which the change in the unperturbed  levels, due to
perturbation, is of the order of the spacing among levels, and
therefore avoided crossings start to appear.
Note that the unperturbed eigenvalues  $k_0^{(nlm)}$ of the sphere are
$(2l+1)$-times 
degenerate with an average spacing $\ave{\Delta k_0} \approx 8/k$.
The derivative with respect to $\epsilon$ of a level $k_0$ of a spherical
billiard 
with an eigenfunction $\Psi_0(\vec{r})$  can be
expressed as an integral over the boundary of the billiard
$ \partial_\epsilon k_0 = 
-(1/2k_0)\int d^2 S f(\vec{r}/r) [\partial_n\Psi_0(\vec{r})]^2$
where $\partial_n$ is a normal derivative w.r.t. boundary.
Then, using the explicit form for $f(\vec{n})$ and $\Psi_0(\vec{r})$, 
we can estimate the level velocities
$\partial_\epsilon k_0 \approx k_0\sqrt{g}/(L+1)$ where $g=48$ is the
symmetry factor. Therefore we obtain:
$$\epsilon_p(k) = \frac{\ave{\Delta k}}{\partial_\epsilon k} \approx
\frac{8(L+1)}{\sqrt{g} k^2}.$$
In fig. 1 we show the dependence of the eigenvalues on the perturbation
parameter $\epsilon$. It is clearly seen that avoided crossings appear
above the perturbative border which, for the parameters of fig. 1
($L=14,k\approx 180.0$), is $\epsilon_p\approx 0.0005$.
Above the perturbative border, namely for $\epsilon > \epsilon_p$ the
quantum diffusion takes place according to
the classical one.Our purpose is to try to understand if and under what conditions
interference terms can suppress quantum diffusion, thus leading to 
localization and deviations from predictions of RMT. As it is known,
quantum localization may 
take place after the Heisenberg time $t_H = 2\pi d{\cal N}/dE = k/36$
at which the quantum motion resolves the discreteness of the energy spectrum.
More precisely, eigenstates are expected to be localized in
the angular momentum coordinates if the Heisenberg time $t_H$
is less than the ergodic time $t_E$ which is the time it takes for the
classical distribution to reach the equilibrium state.
Therefore, transition to
ergodicity
or delocalization takes place at $t_H=t_E$. If we assume, for the time
being, that the classical motion is diffusive with diffusion coefficient
$D\approx \epsilon^2 L^2 E$, then the ergodic time is given by
$t_E\approx \ave{\Delta t} l^2_{\max}/D \approx 8/(3\epsilon^2 L^2 k)$
where $\ave{\Delta t} = 4/(3k)$ is the average time between collisions.
The condition $t_E = t_H$ leads to the estimate for the delocalization
border $\epsilon_l$:
\begin{equation}
\epsilon_l = \frac{4\sqrt{6}}{L k}.
\label{eq:locb}
\end{equation}
However, due to anomalous diffusion, the ergodic time will be longer than
the one given by the above estimate and this
will lead to a delocalization border even larger than 
(\ref{eq:locb}).
For example, for the data of figs. 1-4 ($L=14,k\approx 180$) the
numerically estimated localization border is $\epsilon_l=0.010$.

Billiard's eigenfunctions $\Psi_k(\vec{r})$ can be expanded in terms of
eigenfunctions of a perfectly spherical billiard 
$\braket {\vec r} {nml} = (\sqrt{2}/j_{l+1}(\xi_{ln}))j_l(\xi_{ln}r)Y_{lm}(\vec{n})$, 
$\ket{\Psi_k} = \sum_{nlm}\braket{nlm}{\Psi_k}\ket{nlm}$. 
We define the angular momentum distribution of 
an eigenstate $\Psi_k$ with an eigenvalue $k$ as
$h_k(l) = \frac{1}{\rho_{\rm eq}(l)}\sum\limits_{nm}|\braket{nlm}{\Psi_k}|^2.$
Since we have divided by the classical stationary distribution 
$\rho_{\rm eq}(l)$ then, for completely delocalized, ergodic states,
$h_k(l)$ should approach a constant (apart from fluctuations).
In fig. 2 we show the distributions $h_k(l)$ for three typical eigenstates 
of the billiard at three different values of parameter $\epsilon$.
Eigenstates (a) and (b) are below the delocalization border and are
exponentially localized while (c) is an ergodic extended eigenstate.
However, eigenstates $\braket{nlm}{\Psi}$ 
are extended in the $l_z=m$ variable as we expect as a
consequence of the much faster classical diffusion of $l_z$ w.r.t. $l$.

For a comparison with theoretical predictions, in analogy with the
classical case, we measure the size of eigenfunctions
by the quantity
$\sigma_k = \frac{1}{k}(\bra{\Psi_k}\op{l}^2\ket{\Psi_k}
-\bra{\Psi_k}\op{l}\ket{\Psi_k}^2)^{1/2}.$
We expect that for ergodic (extended) eigenstates, $\sigma_k$ should
agree with the width of the classical steady state angular momentum 
distribution $\sigma_{\rm eq} = 0.2303$.
Instead,for localized states, we expect that 
$ \ave{\sigma_k}\approx\ave{\sigma(l_0,\tau)}_{l_0} $ where $\tau$ is
of the order of Heisenberg time.
In order to suppress fluctuations
the quantum width is averaged over a sufficiently large number of
consecutive eigenstates while the classical width is microcanonically 
averaged over the  initial angular momentum $l_0$.
In order to take into account that, as is empirically known, the
width of eigenstates is approximately one half of the width of steady
state we choose $\tau = t_H/2 = k/72$. The comparison
between $\ave{\sigma_k}$ and $\ave{\sigma(l,t_H/2)}_l$ is shown in
fig. 3 for a fixed energy window. The agreement is surprisingly good.
Moreover we have numerically checked that the above relation also holds
for different energy windows with $k$ ranging between 
$k\approx 120$ and $k\approx 260$ and for fixed $\epsilon=0.003$.

We have also calculated in the different regimes the most commonly studied 
statistical spectral properties, namely the nearest neigbour level spacings 
distribution $P(S)$ and the number variance $\Sigma^2(E)$.
The distribution $P(S)$ has been found to be characterized 
by the power-law level repulsion
$P(S)\propto S^\beta, S\ll 1,$ where $\beta$  smoothly 
increases from zero to one  as a function of $\epsilon$ or $k$ on moving
from the perturbative to the ergodic regime. 
The functional relation 
$\beta(\ave{\sigma_k})$ is still to be systematically investigated.
However, in the perturbative regime $\epsilon < \epsilon_p(k)$, 
$\beta\approx 0$ but the tails of
$P(S)$ decrease much slower than Poissonian
$\exp(-S)$ due to strong degeneracies at $\epsilon=0$.
Moreover the convergence towards GOE distribution with increasing
$\epsilon$ or $k$ is found to be very slow and in agreement with previous 
results on different models [3]. In this respect, the 
$\Sigma^2(E)$ describing the long range correlations has been found to be more sensitive.
In fig. 4 we plot $\Sigma^2(E)$ computed in the window $170 < k < 180$ 
and for $\epsilon = 0.003,0.010,0.016$.
\\\\
In this paper we have shown, on a model of classically chaotic 3D
billiard, the appearence of the quantum dynamical localization phenomenon
which results in the approach of the quantum distribution to a localized,
non ergodic, steady state. This phenomenon, though in a different context,
was discovered long ago \cite{34} and it is related to the existence of
different time scales of classical and quantum motion \cite{35}.

Finally we would like to remark that the 3D nearly spherical billiard
studied here can be a realistic model for a new class of optical
resonators \cite{STO}: in particular the localization in angular
momentum can lead
to an increase in the photons lifetime.

\section*{Figure caption}

\begin{figure}[htbp]
\begin{center}
\leavevmode
\epsfxsize=3.4in
\epsfbox{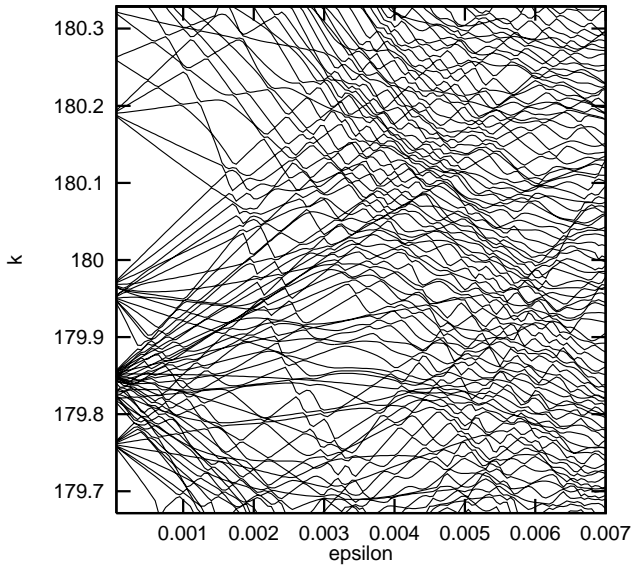}
\end{center}
\caption{
Dependence of quantum eigenvalues on the perturbation parameter 
$\epsilon$ in the window $179.68 < k < 180.32$.
This interval contains approximately 90 levels.} 
\label{fig:1}
\end{figure}

\begin{figure}[htbp]
\begin{center}
\leavevmode
\epsfxsize=3.4in
\epsfbox{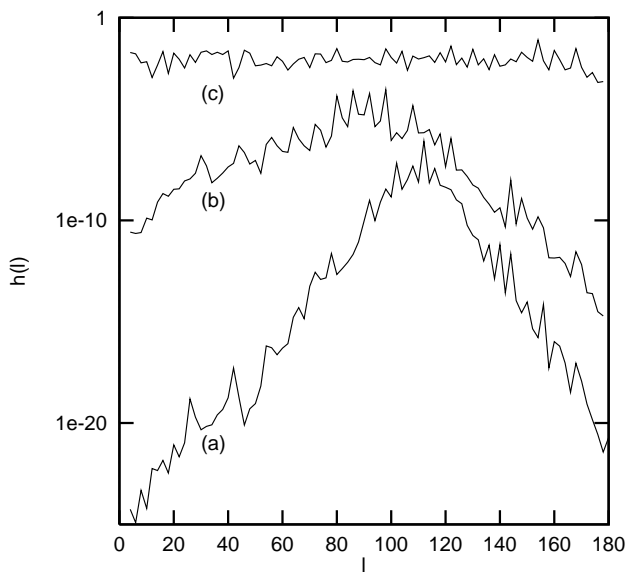}
\end{center}
\caption{
Angular momentum probability distribution $h_k(l)$ of three eigenstates
with $k\approx 180$ and different $\epsilon$:
(a) strongly exponentially localized eigenstate for $\epsilon=0.001$ 
with eigenvalue $k=180.3009$,
(b) localized eigenstate for $\epsilon=0.003$ with eigenvalue $k=179.8013$,
and (c) extended (ergodic) eigenstate for $\epsilon=0.016$ with 
eigenvalue $k=179.8611$. Note that the probability distribution of the state 
(a) is divided by a factor 
$10^6$ and probability for the state (b) by a factor $10^3$ w.r.t. state (c).
} 
\label{fig:2}
\end{figure}

\begin{figure}[htbp]
\begin{center}
\leavevmode
\epsfxsize=3.4in
\epsfbox{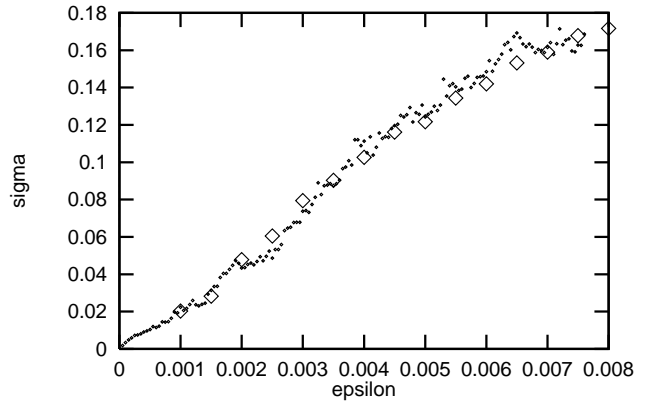}
\end{center}
\caption{
 Comparision of the average width $\ave{\sigma_k}$ of angular momentum 
distribution
of quantum eigenstates  (bullets) and the corresponding widths
 $\ave{\sigma(l_0,t_H/2)}_{l_0}$ of 
classical distributions (diamonds) after half of Heisenberg time
 for different values
of the perturbation parameter $\epsilon$ and fixed $L=14$. The quantum
distributions have been  averaged over $90$ consecutive
states around $k\approx 180$
}
\label{fig:3}
\end{figure}

\begin{figure}[htbp]
\begin{center}
\leavevmode
\epsfxsize=3.4in
\epsfbox{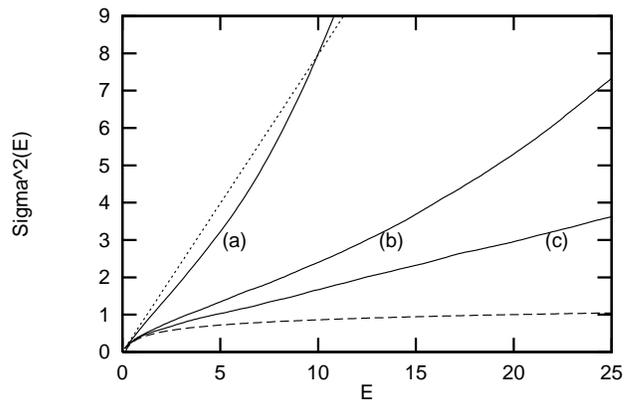}
\end{center}
\caption{
Number variance statistics $\Sigma^2(E)$ for three
spectral stretches in the window $170 < k < 180$ corresponding to
 three different values of the
perturbation parameter (a) $\epsilon=0.003$, (b) $\epsilon = 0.010$,
(c) $\epsilon = 0.016$. Each window contains about $1500$ levels.
The dashed curve is the logarithmic GOE number variance while the dotted
curve is
the linear (Poissonian) $\Sigma^2(E)$ statistics.
The relevant energy scales for this plot are:
mean level spacing $\Delta E = 72\pi/k \approx 1.26$; Thouless energy
$E_c = 1/t_E \approx 3$ for curve (c); the bouncing energy
$E_b = 2/\ave{t} \approx 270$ where $\ave{t}$ is the average bouncing
time. 
} 
\label{fig:4}
\end{figure}

\begin{references}
\bibitem{CGV80}G. Casati, I. Guarneri and F. Valz-Gris, Lettere al
Nuovo Cimento {\bf28},(1980) 279; S.W. McDonald and A.N. Kaufman, Phys.
Rev. Lett. {\bf42},1189(1979).

\bibitem{BGS84}O.Bohigas,M.-J.Giannoni,and C.Schmit, Phys.Rev.Lett.{\bf
52}, 1 (1984); O. Bohigas in {\em Les Houches Lecture Series} {\bf 52},
Eds. M.-J.Giannoni, A.Voros, and J.Zinn-Justin (North-Holland, Amsterdam,
1991).

\bibitem{CCGI}G. Casati, B.V.Chirikov, I. Guarneri and F.M.
Izrailev, Physics Letters A {\bf223},430(1996). 

\bibitem{BCL96}F.Borgonovi, G.Casati, and B.Li,
Phys.Rev.Lett.{\bf 77}, 4744 (1996).

\bibitem{FS97}K.Frahm and D.Shepelyanski,
Phys.Rev.Lett.{\bf78},1440 (1997).

\bibitem{30}H. Alt, H.D. Graf,R. Hofferbert,C. Rangacharyulu,H. Rehfeld, A.
Richter, P. Schardt, and A. Wirzba, phys Rev. E54,2303 (1996).

\bibitem{31}R.L. Weaver, J. Acoust. Soc. Am. 85,1005 (1989).

\bibitem{32}C. Ellegaard, T. Guhr, K. Lindemann, H.Q. Lorensen,J. Nygard, and M.
Oxborrow, Phys. Rev. Lett. 75,1546 (1995).

\bibitem{33}H. Primack and U. Smilansky, Phys. Rev. Lett. 74,4831 (1995).

\bibitem{STO}J.U. Nockel and A.D. Stone, Nature {\bf 385}, 45 (1997).

\bibitem{expl}We have found that the same  anomalous diffusive behaviour is typical of 
the 2D nearly circular billiard
studied in \cite{FS97}. Such behaviour is due to the fact that the
periodic structure of the boundary manifests itself after 
$J^*\approx L$ iterations.

\bibitem{100}For even larger times, the variance $\ave{(l_{zJ}-l_{z0})^2}$
grows considerably slower due to cancellations between terms $\Delta l_{zj}$

\bibitem{VS95}E.Vergini and M.Saraceno, Phys. Rev. E {\bf 52}, 2204 (1995).

\bibitem{P96}T.Prosen, ``Quantization of generic chaotic 3D billiard with smooth
boundary I: Energy level statistics'', preprint;
T.Prosen, ``Quantization of generic chaotic 3D billiard with smooth boundary II: 
structure of high lying eigenstates'', preprint.

\bibitem{34}G. Casati, B.V. Chirikov, J. Ford, and F. M. Izrailev, Lec. Notes
Phys. 93,33 (1979); see also ref [15].

\bibitem{35}G.Casati and B.V. Chirikov, quantun Chaos (Cambridge University
Press, Cambridge, 1995; Physica (Amsterdam)D86, 220 (1995).
\end{references}
\end{document}